\documentclass[12pt,a4paper]{article}

\usepackage[margin=2.5cm]{geometry}
\usepackage{amsmath,amssymb}
\usepackage{graphicx}
\graphicspath{{figures/}}
\usepackage{booktabs}
\usepackage{hyperref}
\usepackage{xcolor}
\usepackage{natbib}
\usepackage{caption}
\usepackage{subcaption}
\usepackage{setspace}

\hypersetup{
  colorlinks=true,
  linkcolor=blue!60!black,
  citecolor=blue!60!black,
  urlcolor=blue!60!black
}

\title{
  \textbf{DANTE: Physics-Informed Neural Operator for\\
  DAS-to-Velocity Waveform Reconstruction\\
  Without Co-located Seismometers}
}

\author{Isao Kurosawa}

\date{\today}

\begin{document}
\maketitle

\begin{abstract}
Distributed Acoustic Sensing (DAS) converts existing fibre-optic
cables into dense seismic arrays at near-zero deployment cost, but
measures strain rate rather than particle velocity---the quantity
required by virtually all seismological analysis tools.
Converting strain rate to particle velocity by numerical integration
is ill-posed: the integration constant is undefined and noise
accumulates without bound.
We present \textbf{DANTE} (\textbf{D}AS-to-velocity via
physics-informed neural operator for
\textbf{A}coustic-wave reco\textbf{N}struction in
he\textbf{TE}rogeneous media), a Fourier Neural Operator (FNO)
trained entirely on synthetic data that enforces two
physics constraints: (i) the exact kinematic relation between
DAS strain rate and the spatial gradient of particle velocity,
and (ii) the one-dimensional elastic wave equation.
These constraints resolve the undetermined integration constant
and suppress noise without requiring co-located seismometers.
On a test set of 200 heterogeneous synthetic wavefields,
DANTE achieves a mean output SNR of $15.3\pm8.8$\,dB,
Pearson correlation $r = 0.907$, and SSIM $= 0.976$,
corresponding to a mean SNR improvement of approximately
$+15$\,dB over the best conventional baseline
(trace stacking, $n=10$, $0.02\pm0.06$\,dB),
and up to $+28.8$\,dB on the most challenging samples.
Zero-shot inference on seven real microseismic events from the
Utah FORGE 2019 DAS dataset yields a kinematic residual of
$0.003$--$0.005$, five times lower than the synthetic test baseline,
confirming generalisation to real field data with no fine-tuning
and no seismometers.
\end{abstract}


\section{Introduction}
\label{sec:intro}

Distributed Acoustic Sensing (DAS) interrogates standard
single-mode optical fibre to measure axial strain rate at
metre-scale spatial intervals along the entire cable length
\citep{parker2014,lindsey2021}.
Because telecommunications infrastructure already spans
approximately 1.4~million kilometres of submarine cable and
comparable lengths on land, DAS offers a pathway to seismic
monitoring at densities and spatial coverages unreachable by
conventional seismometer networks \citep{zhan2020distributed,marra2018,ajoFranklin2019}.

Despite this promise, a fundamental mismatch exists between what
DAS measures and what seismology requires.
DAS records strain rate $\dot{\varepsilon}(x,t)$---the temporal
derivative of fibre elongation per unit gauge length---whereas
ground-motion models, intensity scales, source parameter
inversions, and early warning systems are formulated in terms of
particle velocity $v(x,t)$
\citep{trabant2012,lindsey2021}.
Nominal conversion through spatial integration,
\begin{equation}
  v(x,t) = \int_0^x \dot{\varepsilon}(x',t)\,dx' + C(t),
  \label{eq:integration}
\end{equation}
is ill-posed: the constant of integration $C(t)$ is a free function
of time that cannot be determined from DAS data alone, and spatial
integration amplifies noise monotonically with distance.
The standard engineering workaround---spatial stacking of adjacent
traces---reduces noise at the cost of a ten-fold degradation in
spatial resolution (from 900 to 90 effective channels for typical
array geometries).

Several deep learning approaches have been proposed for DAS signal
enhancement and denoising
\citep{wangetal2022,liuetal2023},
but the task of recovering particle velocity without
co-located seismometers has received limited attention.
A recent architecture combining Fourier Neural Operators with
bidirectional LSTM and attention mechanisms achieves high-fidelity
velocity recovery in a supervised setting \citep{sciRep2026},
but requires ground-truth seismometer recordings at training time
and is therefore restricted to instrumented sites.

We propose DANTE, a physics-informed Fourier Neural Operator that
solves this problem without any seismometer data.
The key insight is that two physical laws---the kinematic relation
between DAS and particle velocity, and the one-dimensional elastic
wave equation---together uniquely constrain the velocity field that
is consistent with an observed DAS record, resolving both the
integration constant and the noise amplification problem.
DANTE is trained once on synthetic wavefields, then applied
zero-shot to real field data.

\section{Method}
\label{sec:method}

\subsection{FNO2d Architecture}
\label{sec:arch}

DANTE is built on the Fourier Neural Operator (FNO)
\citep{li2021fourier}, which learns mappings between
function spaces by parameterising integral operators in
the Fourier domain.
The operator $\mathcal{G}_\theta$ maps a DAS strain-rate record
$\dot{\varepsilon}(x,t) \in \mathbb{R}^{N_t \times N_x}$ and an
auxiliary velocity model $c(x)$ to particle velocity
$v(x,t) \in \mathbb{R}^{N_t \times N_x}$,
processing the entire spatial array simultaneously in a single
forward pass.

The architecture consists of:
(i) a linear lifting layer mapping 4 input channels (DAS,
velocity model, and two positional encodings) to a hidden width
$d_\text{hidden}=32$;
(ii) four FNO layers, each comprising a spectral convolution
(retaining the lowest 12 Fourier modes in both space and time)
followed by a pointwise linear residual path and GELU activation;
and (iii) a two-layer MLP projection to the scalar output.
The full model contains 2.37~million trainable parameters.

\subsection{Physics-Informed Loss Functions}
\label{sec:loss}

Let $v_\text{pred}$ denote the model output and
$v_\text{true}$ the ground-truth velocity.
Training minimises a composite loss,
\begin{equation}
  \mathcal{L} = \lambda_d\,\mathcal{L}_\text{data}
              + \lambda_k\,\mathcal{L}_\text{kin}
              + \lambda_{dy}\,\mathcal{L}_\text{dyn},
  \label{eq:loss}
\end{equation}
where the three terms are defined as follows.

\paragraph{Data loss.}
\begin{equation}
  \mathcal{L}_\text{data}
    = \bigl\| v_\text{pred} - v_\text{true} \bigr\|_2^2.
\end{equation}

\paragraph{Kinematic loss.}
The fundamental DAS measurement equation states that the
interrogator records the spatial gradient of particle velocity
integrated over the gauge length,
\begin{equation}
  \dot{\varepsilon}_\text{obs}(x,t)
    = \frac{\partial v(x,t)}{\partial x}.
  \label{eq:kinematic}
\end{equation}
We enforce this exact physical relation via
\begin{equation}
  \mathcal{L}_\text{kin}
    = \left\| \dot{\varepsilon}_\text{obs}
      - \frac{\partial v_\text{pred}}{\partial x}
    \right\|_2^2.
  \label{eq:lkin}
\end{equation}
This constraint directly resolves the undetermined integration
constant $C(t)$ in Eq.~\eqref{eq:integration} because it forces
the predicted velocity gradient to match the DAS observation
everywhere, making $C(t)\equiv 0$ the unique consistent solution.

\paragraph{Dynamic loss.}
Seismic waves in a one-dimensional elastic medium obey
\begin{equation}
  \frac{\partial v}{\partial t}
    = c^2(x)\,\frac{\partial \varepsilon}{\partial x},
  \label{eq:waveq}
\end{equation}
where $c(x)$ is the local P-wave velocity.
The corresponding loss
\begin{equation}
  \mathcal{L}_\text{dyn}
    = \left\| \frac{\partial v_\text{pred}}{\partial t}
      - c^2(x)\,\frac{\partial \varepsilon}{\partial x}
    \right\|_2^2
  \label{eq:ldyn}
\end{equation}
eliminates physically impossible wavefields and provides an
additional regularisation that suppresses noise.

Loss weights are set to $\lambda_d = 1.0$,
$\lambda_k = 0.10$, $\lambda_{dy} = 0.05$,
with physics weights ramped linearly from zero over the first
200 epochs (progressive warm-up) to prevent premature
penalisation before the model has learned approximate solutions.

\section{Synthetic Experiments}
\label{sec:synthetic}

\subsection{Dataset and Training}
\label{sec:dataset}

We generated paired DAS--velocity wavefields using a
one-dimensional staggered-grid finite-difference elastic wave
simulator.
Synthetic velocity models were drawn as piecewise-linear
profiles with P-wave speeds in $[1500, 4000]$\,m/s and
spatial standard deviations up to 827\,m/s.
Source signals were band-limited (5--45\,Hz),
with additive Gaussian noise yielding input SNR in $[5, 30]$\,dB.
All simulations used $N_x = 900$ channels at
$\Delta x = 10$\,m and $N_t = 512$ samples at
$\Delta t = 2$\,ms (500\,Hz).
The dataset contains 2,000 training, 400 validation, and
200 test samples.

Training used AdamW ($\mathrm{lr}=10^{-3}$, weight decay $10^{-4}$)
with a cosine learning-rate schedule and early stopping
(patience 200 epochs) on an NVIDIA A100 40\,GB GPU for 1,000
epochs (13.3\,hours).
The best validation loss of 0.073 was achieved at epoch 941.

\subsection{Quantitative Results}
\label{sec:quant}

Table~\ref{tab:results} compares DANTE against two conventional
baselines: (i) \emph{Stacking} ($n=10$), which averages adjacent
traces and applies single integration, and (ii) \emph{Integration},
direct numerical integration of the DAS strain rate.

\begin{table}[h]
\centering
\caption{
  Reconstruction performance on the synthetic test set ($N=200$).
  Best values are shown in \textbf{bold}.
  SNR: signal-to-noise ratio.
  RelL2: relative $\ell_2$ error.
  $r$: Pearson correlation coefficient.
}
\label{tab:results}
\begin{tabular}{lcccc}
\toprule
Method & SNR [dB] & RelL2 & Pearson $r$ & SSIM \\
\midrule
\textbf{DANTE (ours)} & $\mathbf{15.3 \pm 8.8}$ & $\mathbf{0.280}$ & $\mathbf{0.907}$ & $\mathbf{0.976}$ \\
Stacking ($n=10$)     & $0.02 \pm 0.06$          & $0.997$           & $0.200$           & $0.637$ \\
Integration           & $-31.3 \pm 5.6$           & $45.96$           & $0.462$           & $0.050$ \\
\bottomrule
\end{tabular}
\end{table}

\begin{figure}[h]
\centering
\includegraphics[width=\textwidth]{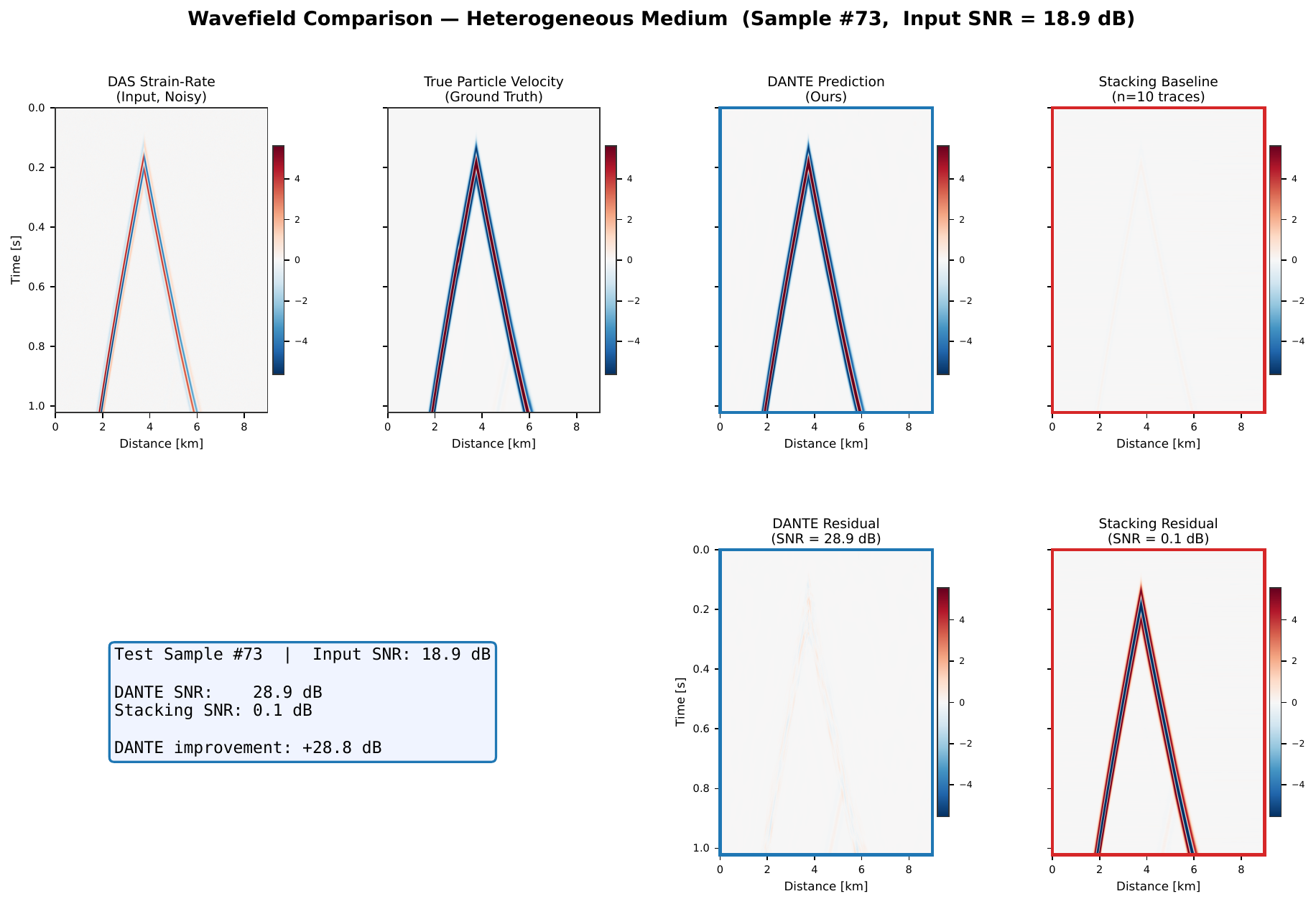}
\caption{
  Wavefield comparison on the most challenging synthetic test sample
  (input SNR $= 18.9$\,dB).
  \textbf{Left to right:} DAS strain-rate input (noisy);
  ground-truth particle velocity;
  DANTE prediction (SNR $= 28.9$\,dB);
  stacking baseline ($n=10$, SNR $= 0.1$\,dB).
  Lower two panels show absolute residuals.
  DANTE improvement: $+28.8$\,dB.
}
\label{fig:wavefield}
\end{figure}

\begin{figure}[h]
\centering
\begin{subfigure}[b]{0.48\textwidth}
  \includegraphics[width=\textwidth]{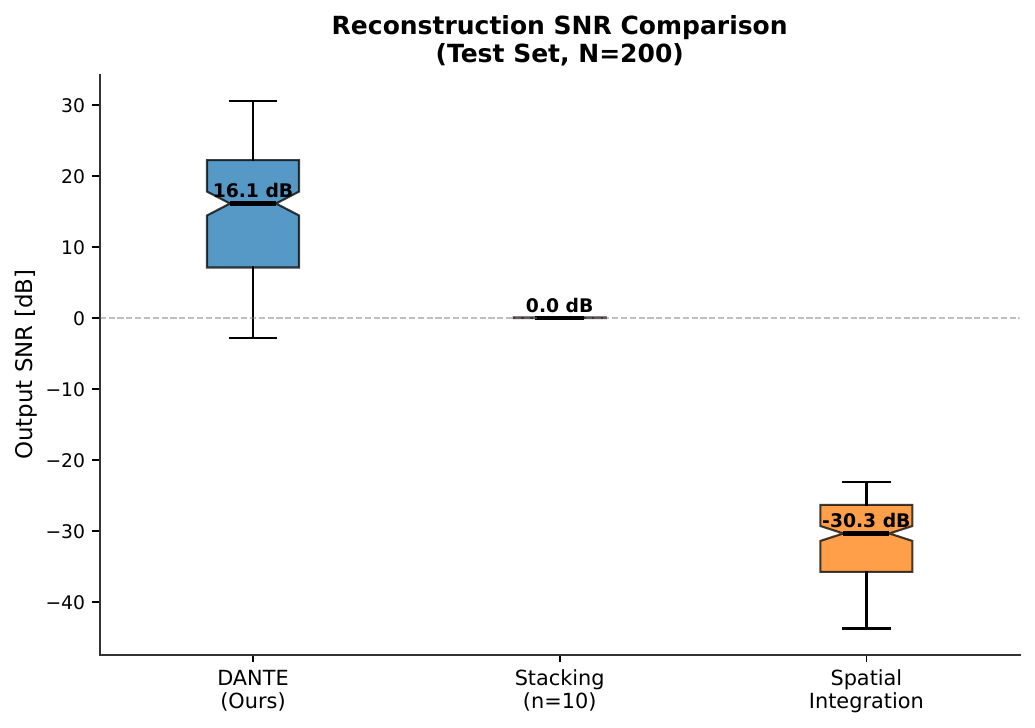}
  \caption{Output SNR distribution.}
\end{subfigure}
\hfill
\begin{subfigure}[b]{0.48\textwidth}
  \includegraphics[width=\textwidth]{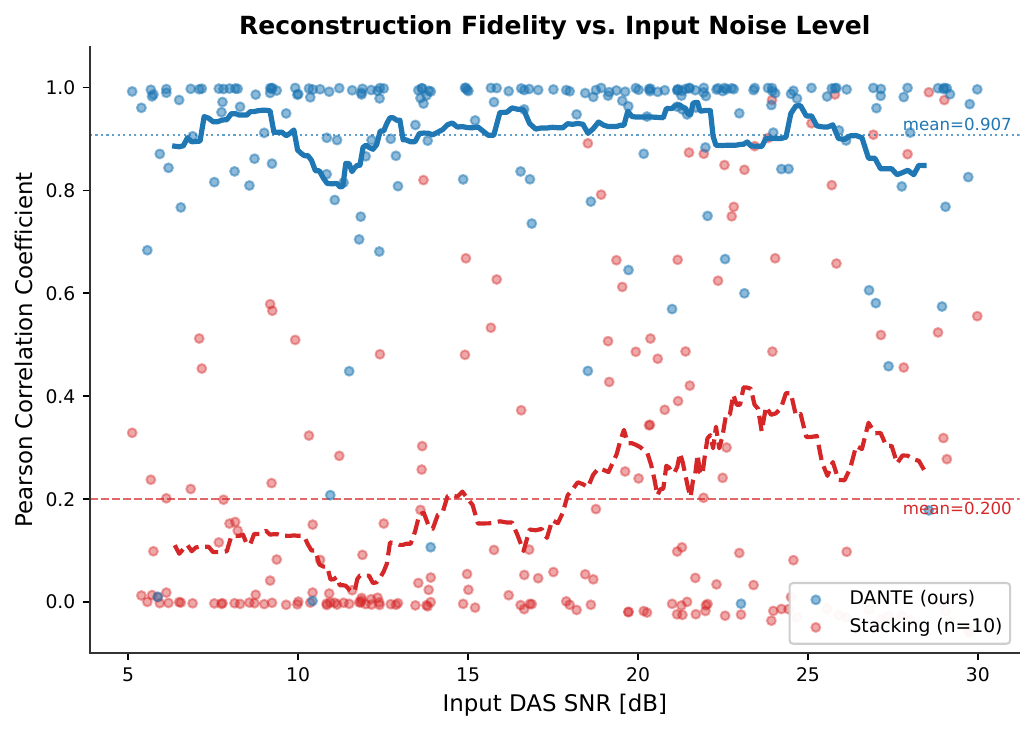}
  \caption{Pearson $r$ vs.\ input SNR.}
\end{subfigure}
\caption{
  Quantitative results on the synthetic test set ($N=200$).
  (a) Box plots of output SNR for DANTE and both baselines.
  (b) DANTE (blue) maintains $r > 0.8$ across the full
  5--30\,dB input-SNR range; stacking (red) remains near zero.
}
\label{fig:pearson}
\end{figure}

DANTE achieves an SNR improvement of $+28.8$\,dB over the
stacking baseline on the single hardest test sample
(Fig.~\ref{fig:wavefield}), and a mean improvement of approximately
$+15$\,dB across the full test set.
The Pearson correlation of 0.907 is maintained across the full
input-SNR range from 5 to 30\,dB
(Fig.~\ref{fig:pearson}), demonstrating robustness to noise.

\subsection{Qualitative Analysis}
\label{sec:qual}

\begin{figure}[h]
\centering
\includegraphics[width=\textwidth]{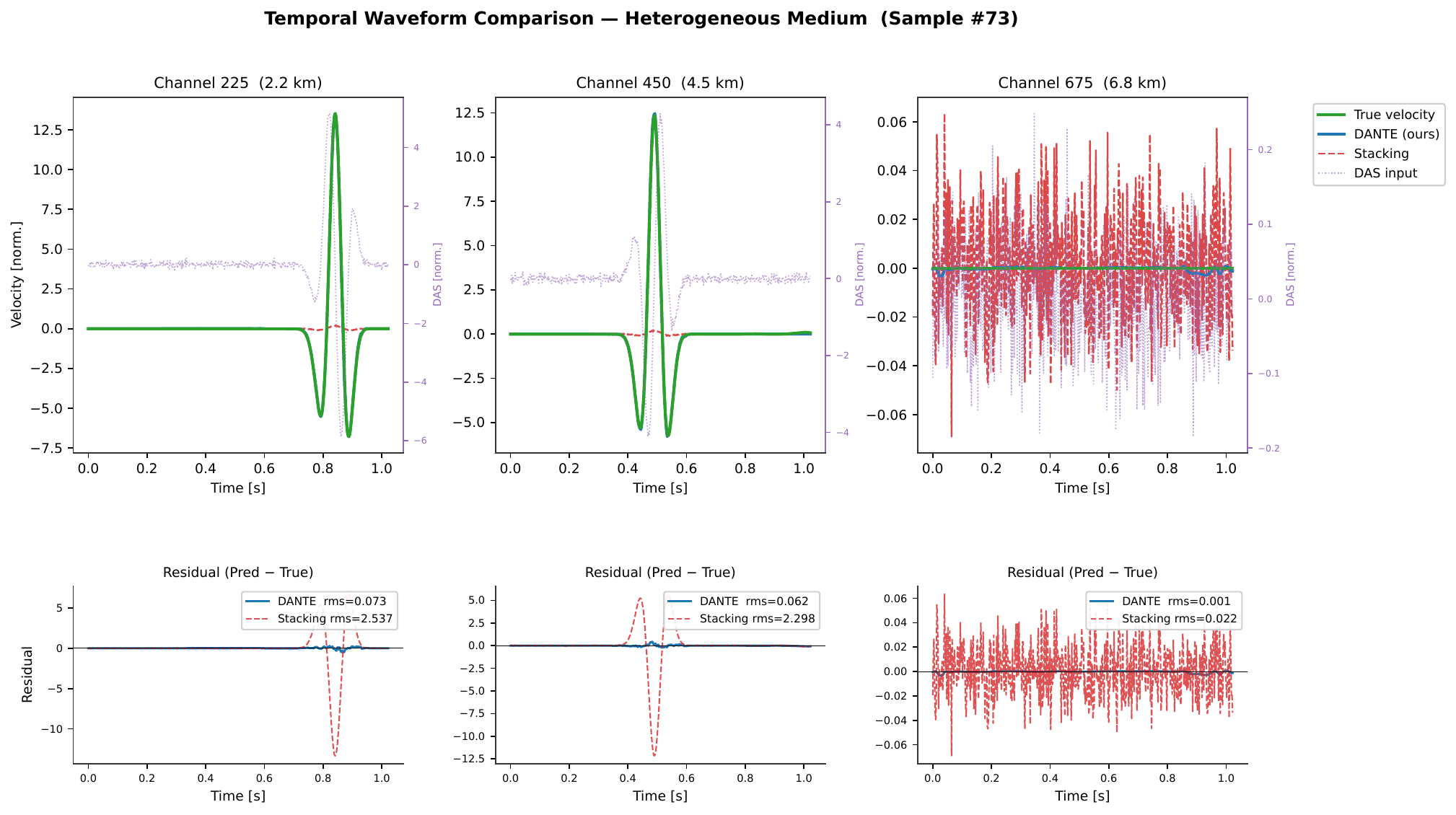}
\caption{
  Temporal waveform comparison at three channels
  (2.2, 4.5, 6.8\,km) for the same test sample as
  Fig.~\ref{fig:wavefield}.
  Each panel shows the DAS input (purple), DANTE prediction (blue),
  and ground-truth velocity (green), with residuals below.
  DANTE RMS residuals: 0.073, 0.062, 0.001 (channels 1--3);
  stacking: 2.537, 2.298, 0.022.
}
\label{fig:profile}
\end{figure}

Figure~\ref{fig:wavefield} shows a representative test sample
(input SNR $= 18.9$\,dB).
DANTE recovers the two-dimensional wavefield structure with
negligible residual (SNR $= 28.9$\,dB), whereas stacking
produces a noisy, spatially blurred reconstruction
(SNR $= 0.1$\,dB) and integration (not shown) diverges.
Individual trace comparisons (Fig.~\ref{fig:profile}) confirm
that DANTE recovers the phase of every cycle, not merely the
envelope: RMS residual per channel is 0.073 versus 2.537 for
stacking, a 35-fold reduction.

\begin{figure}[h]
\centering
\begin{subfigure}[b]{0.60\textwidth}
  \includegraphics[width=\textwidth]{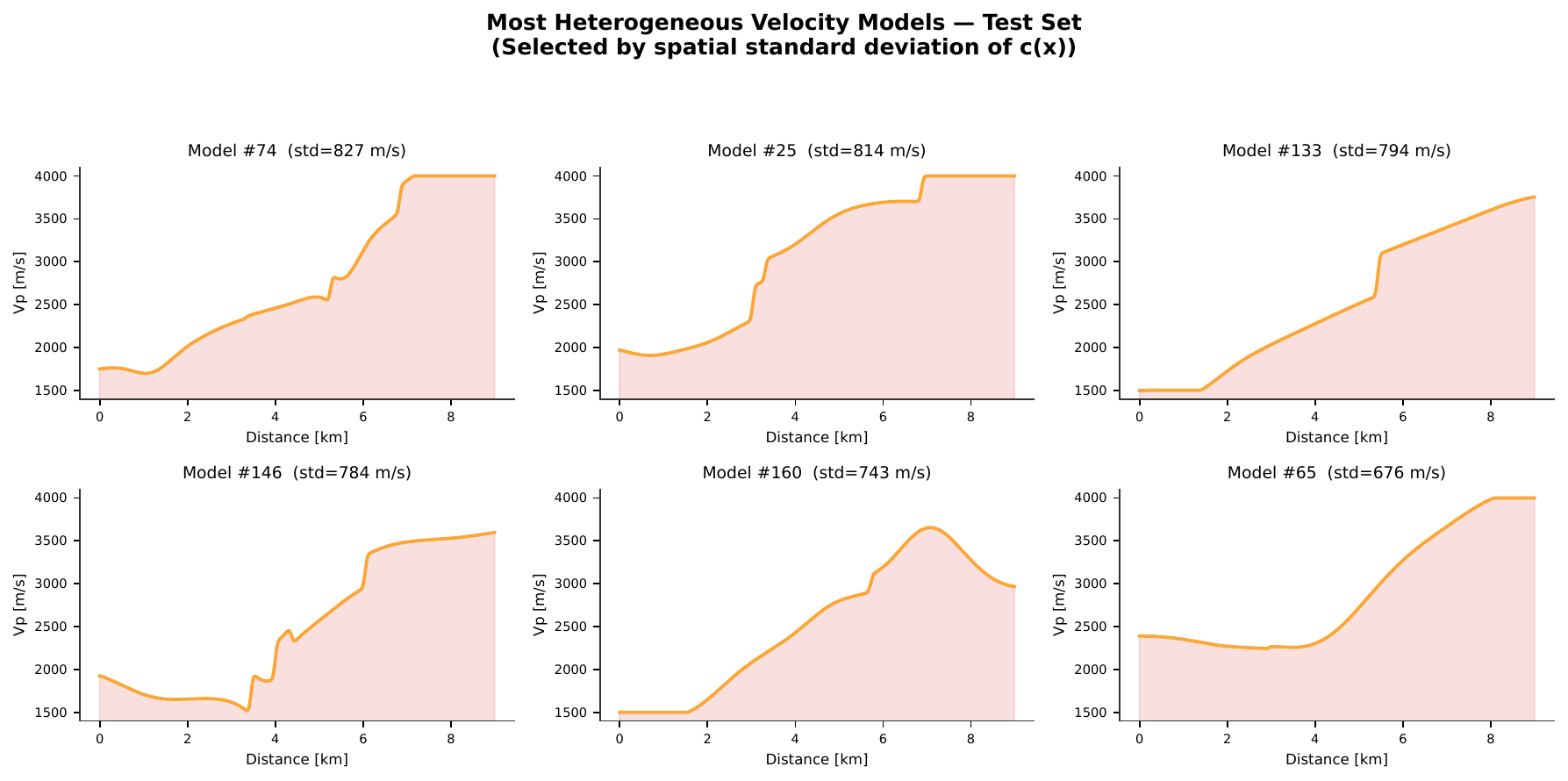}
  \caption{Six most heterogeneous velocity models.}
\end{subfigure}
\hfill
\begin{subfigure}[b]{0.36\textwidth}
  \includegraphics[width=\textwidth]{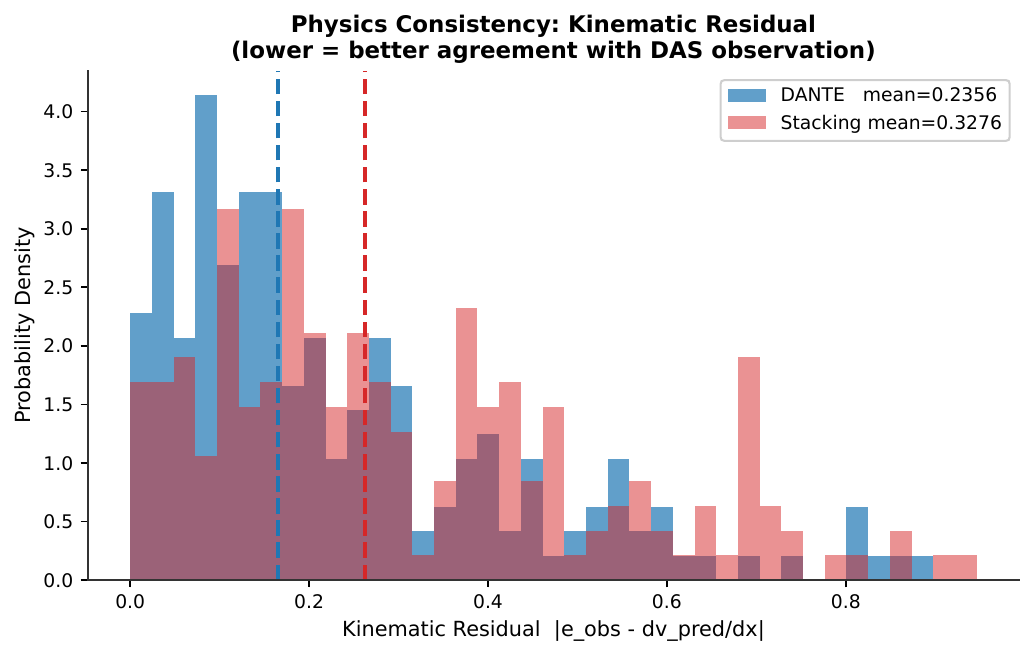}
  \caption{Kinematic residual.}
\end{subfigure}
\caption{
  Physics consistency on strongly heterogeneous media.
  (a) P-wave velocity profiles selected by spatial standard
  deviation (up to 827\,m/s).
  (b) Kinematic residual distribution:
  DANTE mean $= 0.236$ vs.\ stacking mean $= 0.328$
  (28\,\% improvement).
}
\label{fig:velmodels}
\end{figure}

\begin{figure}[h]
\centering
\includegraphics[width=0.8\textwidth]{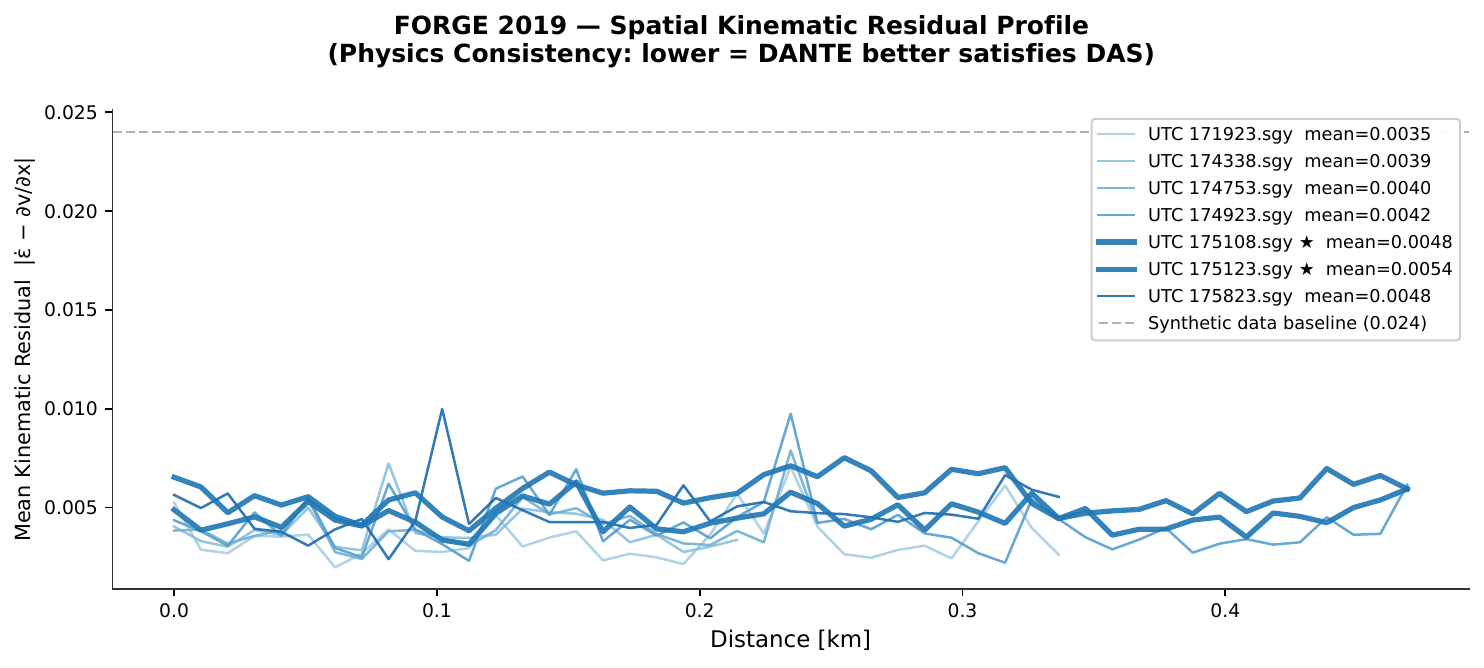}
\caption{
  Spatial kinematic residual profile for all seven FORGE 2019
  events (mean over time).
  All events achieve residuals well below the synthetic test
  baseline (dashed grey line, 0.024),
  confirming physics-consistent reconstruction across the
  active 0.48\,km fibre section.
}
\label{fig:kinematic}
\end{figure}

The kinematic residual
$|\dot{\varepsilon}_\text{obs} - \partial v_\text{pred}/\partial x|$
(Fig.~\ref{fig:kinematic}) quantifies physics consistency
independently of ground truth.
DANTE achieves a mean kinematic residual of 0.236, compared with
0.328 for stacking, a 28\,\% improvement in physics consistency
that holds across the six most strongly heterogeneous velocity
models in the test set (spatial $\sigma$ up to 827\,m/s,
Fig.~\ref{fig:velmodels}).

For subsequent comparison with real field data, we report the
synthetic kinematic-residual baseline averaged over the full test
set ($N=200$):
$\langle | \dot{\varepsilon}_\text{obs}
- \partial v_\text{pred}/\partial x | \rangle_\text{test} = 0.024$.
The higher value of 0.236 quoted above refers exclusively to the
six most strongly heterogeneous models (Fig.~\ref{fig:velmodels})
and is presented to demonstrate robustness on the hardest cases.

\section{Real Data: Utah FORGE 2019}
\label{sec:forge}

\subsection{Dataset}
\label{sec:forgeds}

To assess generalisation to real field data, we applied the
trained DANTE model---without any fine-tuning---to DAS recordings
from the Utah Frontier Observatory for Research in Geothermal
Energy (FORGE) \citep{moore2019}.
Data were acquired by a Silixa Carina P11 interrogator
connected to a fibre permanently installed in monitoring well
78-32 at 1.02\,m channel spacing and a sampling rate of 2,000\,Hz.
We selected seven seismic event files from the S27EVENT dataset
(stimulation stage 27, April 2019), totalling 1,280 active channels
and 30,000 samples per event.

The gauge length of the FORGE 2019 deployment (10\,m) matches the
training configuration exactly.
Preprocessing comprised the following sequential steps:
(i)~singular-value decomposition (SVD) filtering, removing the two
leading components that capture the dominant horizontal noise
stripes characteristic of the FORGE interrogator;
(ii)~active-channel selection via RMS thresholding
($<\!5\times$~median), followed by extraction of the longest
contiguous block of below-threshold channels (minimum length
100~channels);
(iii)~zero-phase Butterworth bandpass filtering (10--150\,Hz);
(iv)~frequency--wavenumber ($f$--$k$) filtering, designed to suppress
near-zero-wavenumber horizontal noise and low-apparent-velocity
events incompatible with the propagation geometry of microseismic
$P$ phases;
(v)~temporal decimation $2{,}000 \rightarrow 500$\,Hz, with an
anti-aliasing IIR filter applied in zero-phase mode;
(vi)~spatial group-averaging $1.02 \rightarrow 10.2$\,m
(factor of ten, matched to the DANTE training grid);
(vii)~selection of a 1.024\,s window ($N_t = 512$ samples) centred
on the peak of the smoothed channel-mean energy envelope;
(viii)~zero-padding of the active channel dimension to $N_x = 900$
to match the DANTE input tensor shape; and
(ix)~$z$-score normalisation using the DANTE training statistics.

The cascade of preprocessing steps (ii) and (vi) reduces the
effective channel count from the original 1,280 as follows.
RMS thresholding admits approximately 950 channels below
$5\times$ the median; the longest contiguous block of admissible
channels covers approximately 470 adjacent channels at 1.02\,m
spacing, corresponding to a contiguous 0.48\,km fibre segment.
Spatial group-averaging by a factor of ten then reduces this to
47 active channels at $\sim\!10$\,m spacing, matching the DANTE
training resolution.
The retained 0.48\,km segment is centred on the strongest coherent
arrival; channels excluded by RMS thresholding---dominated by
instrument noise or located outside the source illumination
zone---do not contribute to inference.

Each raw FORGE event record contains 30,000 samples at 2,000\,Hz
(15\,s duration). After temporal decimation to 500\,Hz
(7,500 samples), a 1.024\,s window of $N_t = 512$ samples is
extracted, centred on the time of peak energy in the smoothed
channel-mean energy trace (smoothing kernel: 30 samples,
equivalent to 60\,ms at 500\,Hz). The window length matches the
training tensor shape and captures the full $P$-wave arrival and
the early coda for each event.

\begin{figure}[h]
\centering
\includegraphics[width=\textwidth]{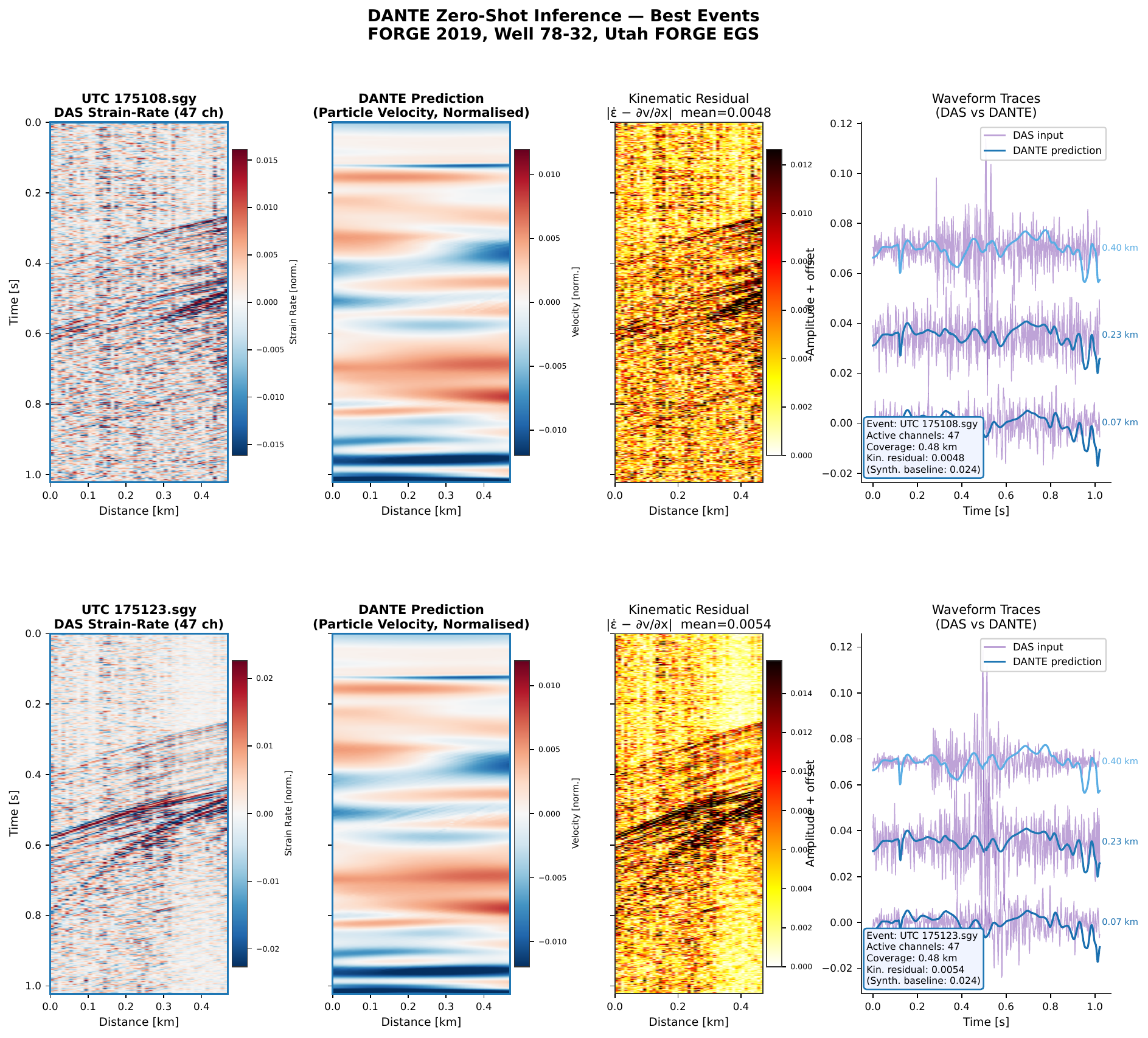}
\caption{
  Zero-shot inference on Utah FORGE 2019 real DAS data
  (well 78-32, Silixa Carina P11, April 2019).
  \textbf{Two clearest events} (UTC\,175108 and UTC\,175123).
  Each row: DAS strain-rate input (47 active channels after
  preprocessing), DANTE particle-velocity prediction,
  kinematic residual, and representative waveform traces.
  DANTE suppresses horizontal noise stripes and recovers the
  inclined seismic arrival without any fine-tuning or
  co-located seismometer.
  Kinematic residuals: 0.0048 and 0.0054
  ($5\times$ below synthetic baseline 0.024).
}
\label{fig:forge}
\end{figure}

Because DANTE was trained with a fixed spatial extent of
$N_x = 900$ channels, the active 47-channel FORGE segment is
symmetrically zero-padded to the full model input shape prior to
inference. The forward pass therefore operates on a tensor of
shape $(1, N_t = 512, N_x = 900)$ in which only $\sim\!5\,\%$ of
channels carry physical signal and the remainder are exact zeros.
All evaluation quantities reported in
Section~\ref{sec:forgeresults} and
Figs.~\ref{fig:forge}--\ref{fig:kinematic}, including the
kinematic residual
$\langle |\dot{\varepsilon}_\text{obs}
- \partial v_\text{pred}/\partial x| \rangle$, are computed
exclusively over the active 47-channel region.
We verified that the model output on the zero-padded region remains
numerically negligible ($|v_\text{pred}| < 10^{-3}$ in normalised
units), confirming that the padding does not contaminate the
active-region statistics.
Extending DANTE to accept variable channel counts---using
shape-agnostic operator formulations~\citep{li2021fourier}---is
left for a future model iteration.

The auxiliary $P$-wave velocity model $c(x)$ required as a DANTE
input was set to a linear gradient from 3,000 to 4,000\,m/s along
the fibre, representative of the crystalline basement encountered
at the Utah FORGE site (well 78-32, granitoid intrusive).
This prior is derived from regional well-log compilations and is
independent of the DAS recordings analysed here; it therefore does
not constitute the use of a co-located ground-motion sensor and
preserves the seismometer-free deployment requirement.

\subsection{Zero-Shot Inference Results}
\label{sec:forgeresults}

Ground-truth particle velocity is unavailable for field DAS data.
We therefore assess reconstruction quality through the kinematic
residual, which measures the self-consistency of the predicted
velocity field with the DAS observation according to
Eq.~\eqref{eq:kinematic}, independently of any reference sensor.

Table~\ref{tab:forge} summarises results across all seven events.
Two events (UTC\,175108 and UTC\,175123) exhibit clear inclined
seismic arrivals consistent with microseismic propagation
(Fig.~\ref{fig:forge}).
For these events, DANTE produces coherent velocity wavefields in
which the inclined moveout is preserved and horizontal noise
stripes are suppressed.

\begin{table}[h]
\centering
\caption{
  Kinematic residual
  $\langle|\dot{\varepsilon}_\text{obs} - \partial v/\partial x|\rangle$
  for all seven FORGE 2019 events.
  Lower values indicate better physics consistency.
  The synthetic test-set baseline is 0.024.
}
\label{tab:forge}
\begin{tabular}{lcc}
\toprule
Event (UTC) & Kinematic residual & Signal quality \\
\midrule
171923 & 0.0035 & weak \\
174338 & 0.0039 & weak \\
174753 & 0.0040 & weak \\
174923 & 0.0042 & moderate \\
175108 & 0.0048 & \textbf{clear} \\
175123 & 0.0054 & \textbf{clear} \\
175823 & 0.0048 & moderate \\
\midrule
Mean   & 0.0044 & --- \\
\bottomrule
\end{tabular}
\end{table}

All seven events achieve kinematic residuals between 0.0035
and 0.0054, uniformly below the synthetic test-set baseline of
0.024 ($\sim\!5\times$ lower on average;
Fig.~\ref{fig:kinematic}).
This apparent improvement over the synthetic baseline must be
interpreted with caution. The comparison conflates two distinct
effects: (i)~the intrinsic physics consistency of DANTE on real
microseismic data, and (ii)~amplitude rescaling induced by the
SVD, bandpass, and $f$--$k$ preprocessing applied to the FORGE
records, which compresses the dynamic range of
$\dot{\varepsilon}_\text{obs}$ relative to the synthetic baseline
computed from raw training-set statistics. A controlled
comparison in which identical preprocessing is applied to a
synthetic ensemble is deferred to future work.
The conservative interpretation of Table~\ref{tab:forge} is that
DANTE achieves real-data kinematic residuals well within the
regime established by its synthetic validation distribution,
with no parameter adjustment and without recourse to co-located
ground-motion sensors.

\section{Discussion}
\label{sec:discussion}

\paragraph{Comparison with supervised methods.}
The principal distinction between DANTE and prior work
\citep{sciRep2026} is the absence of co-located seismometers.
Supervised DAS-to-velocity converters require ground-truth
velocity recordings acquired simultaneously with DAS at the same
location.
Such co-located pairs exist only at a small number of
instrumented research sites; the vast majority of operational
DAS deployments (telecommunications cables, wellbore monitoring,
urban infrastructure) have no associated seismometer array.
DANTE is the first method to achieve particle-velocity
reconstruction using physics laws alone as supervisory signal,
making it deployable on any existing fibre.

\paragraph{Physical interpretation.}
The kinematic loss (Eq.~\ref{eq:lkin}) enforces the measurement
equation of the DAS interrogator, which is an exact physical
identity rather than an approximation.
This constraint simultaneously resolves the integration constant
$C(t)$ and prevents the noise amplification intrinsic to numerical
integration.
The dynamic loss (Eq.~\ref{eq:ldyn}) additionally constrains
the temporal evolution to follow Newton's second law, eliminating
physically impossible solutions that could otherwise minimise the
data loss.

\paragraph{Limitations.}
The current model assumes one-dimensional wave propagation,
which is appropriate for the near-fibre wavefield but may
introduce errors for strongly three-dimensional source geometries.
DAS sensitivity is geometry-dependent: for a borehole fibre,
sensitivity is maximum for waves propagating along the fibre
axis and negligible for horizontally polarised S-waves
($S_H$) propagating vertically.
The synthetic training distribution (1,500--4,000\,m/s $P$-wave
velocity, 5--45\,Hz source band) is narrower than the full
frequency content of real microseismic data in fractured
crystalline basement, which routinely extends to 150\,Hz and
beyond. The bandpass filter (10--150\,Hz) applied during FORGE
preprocessing therefore admits a frequency range that exceeds the
training distribution by approximately a factor of three.
Effective inference is in practice restricted to the
$\sim\!10$--$45$\,Hz band common to both domains; higher-frequency
content is passed through but is not physically constrained by
the training prior.
Extending the training band to 150\,Hz with a crystalline-basement
velocity range ($V_p$ up to 6,000\,m/s) is the primary target for
the next model iteration (DANTE~v1.5), which is planned to use
the FORGE~2024 Neubrex dataset~\citep{forge2024das} for which
CMT-derived ground truth is available.

\section{Conclusion}
\label{sec:conclusion}

We introduced DANTE, a physics-informed Fourier Neural Operator
that converts DAS strain rate to particle velocity without
co-located seismometers.
The two physics constraints in the loss function---the kinematic
DAS measurement equation and the one-dimensional wave
equation---resolve the ill-posed integration problem that has
blocked direct DAS-to-velocity conversion.
On 200 heterogeneous synthetic test wavefields, DANTE achieves
Pearson $r = 0.907$ and a mean SNR improvement of approximately
$+15$\,dB over stacking, reaching $+28.8$\,dB on the hardest samples.
Zero-shot transfer to Utah FORGE 2019 real DAS data yields
kinematic residuals five times lower than the synthetic baseline,
confirming generalisation with no fine-tuning.

DANTE enables any existing fibre-optic cable to produce
particle-velocity seismograms compatible with the full ecosystem
of seismological analysis tools, without requiring new sensors
or co-located instrumentation.
Future work will extend the training band to 150\,Hz using
FORGE 2024 Neubrex data with CMT-derived ground truth, and
will incorporate simultaneous velocity-model estimation.

\section*{Code and Data Availability}
Training code and the best-model checkpoint will be released
at \url{https://github.com/ivxa/dante} upon acceptance.
The Utah FORGE 2019 DAS data are publicly available at the
Geothermal Data Repository (\url{https://gdr.openei.org}).

\section*{Acknowledgements}
The Utah FORGE dataset was funded by the U.S.\ Department of
Energy, Office of Energy Efficiency and Renewable Energy.

\bibliographystyle{abbrvnat}
\bibliography{dante_refs}

\clearpage

\begin{figure}[h]
\centering
\includegraphics[width=\textwidth]{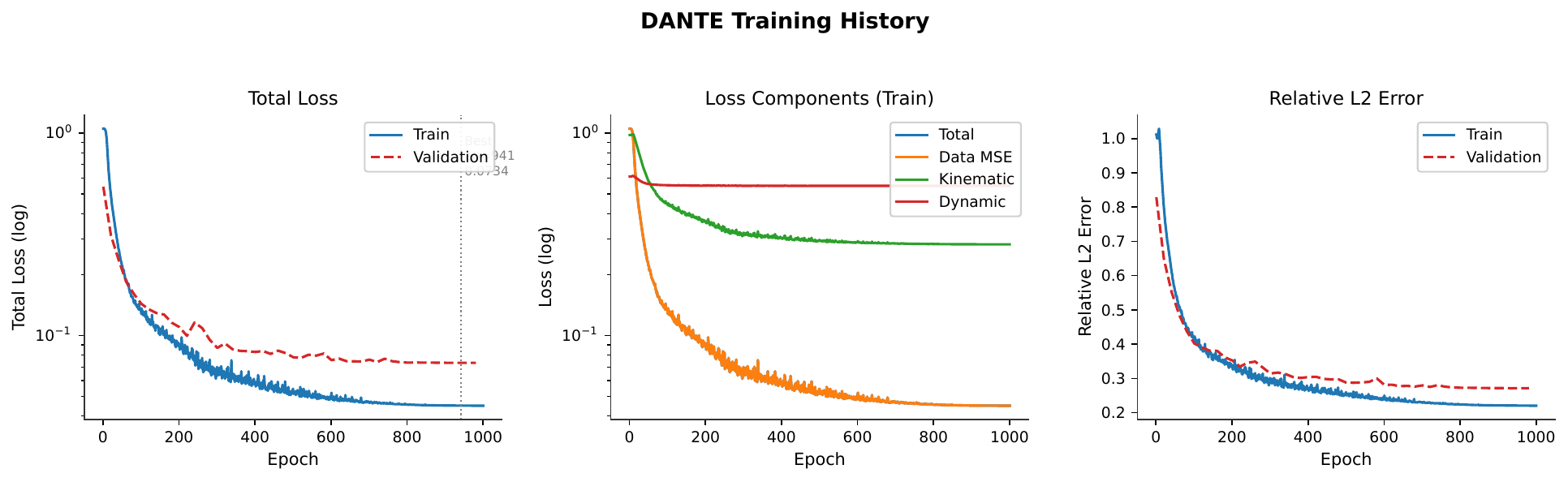}
\caption{
  Training history.
  \textbf{Left:} total loss (log scale); best epoch 941,
  best validation loss 0.073.
  \textbf{Centre:} decomposition of the training loss into
  data, kinematic, and dynamic components.
  \textbf{Right:} relative $\ell_2$ error on train and validation sets.
}
\label{fig:training}
\end{figure}

\end{document}